\documentclass[trackchanges,twocolumn]{aastex701}
\usepackage{amsmath}

\begin{document}

\title{A Calorimetric Framework for Energy Partition in GW170817/GRB 170817A}

\author{Rupam Samanta}
\affiliation{School of Physics, Indian Institute of Science Education and Research (IISER) Thiruvananthapuram, 695551, India}
\email{rupam24@iisertvm.ac.in}

\author{Shabnam Iyyani}
\affiliation{School of Physics, Indian Institute of Science Education and Research (IISER) Thiruvananthapuram, 695551, India}
\affiliation{Centre for High Performance Computing, Indian Institute of Science Education and Research (IISER) Thiruvananthapuram, 695551, India }
\email{shabnam@iisertvm.ac.in}

\author{Surya Vamshi Allada}

\affiliation{School of Physics, Indian Institute of Science Education and Research (IISER) Thiruvananthapuram, 695551, India}

\affiliation{Astronomical Institute, Tohoku University, Sendai 980-0845, Japan}

\email{allada.surya@astr.tohoku.ac.jp}

\correspondingauthor{\\
$^{\dagger}$Rupam Samanta \href{mailto:rupam24@iisertvm.ac.in}{rupam24@iisertvm.ac.in}\\
$^{\ddagger}$Shabnam Iyyani: \href{mailto:shabnam@iisertvm.ac.in}{shabnam@iisertvm.ac.in}\\
}

\begin{abstract}

The binary neutron star merger GW170817 and its electromagnetic counterpart GRB 170817A provided the first joint detection of gravitational waves and a gamma-ray burst. Owing to its exceptional observational coverage, it remains the only binary neutron star merger for 
which the dominant energy channels can be observationally constrained. We present the first observationally constrained calorimetric framework for GW170817/GRB 170817A, treating the system as a closed energy reservoir 
anchored to the ADM-effective energy at the onset of observable inspiral. Combining gravitational-wave and electromagnetic observations with baryonic-mass conservation yields a quantitative merger energy budget. We find that gravitational waves carry away $\sim1.6\%$ of the total energy, 
while the complete electromagnetic output, including prompt emission, kilonova, and afterglow, contributes less than $0.22\%$. With $\sim5$ -- $10\%$ retained in the disk and dynamical ejecta, more than $88\%$ remains in the compact remnant and neutrino reservoir. Global energy closure and baryonic-mass bookkeeping constrain the compact-remnant mass to $2.352$ -- $2.403 \,M_\odot$ (central 68\% interval), adopting the EOS-insensitive relation for rotating neutron stars, and imply a calorimetric upper limit on the neutrino energy of $(1.77$ -- $2.35)\times10^{53} \, \mathrm{erg}$ (central 68\% interval), consistent with current neutrino non-detections. We show that the inferred partition 
can estimate prompt electromagnetic output in mergers without detected gamma-ray bursts and constrain the inclination angle--luminosity distance degeneracy. Extending this analysis to a larger sample can provide an empirical basis for predicting electromagnetic outputs and detection prospects of future mergers.

\end{abstract}


\section{Introduction}
The binary neutron star (BNS) merger GW170817 \citep{Abbott2017GW,Abbott2019GWTC1}
marked the beginning of multi-messenger gravitational-wave astronomy. The joint observation of gravitational waves and the short gamma-ray burst GRB~170817A \citep{Goldstein2017Fermi,Savchenko2017, vonKienlin2017GCN}, followed by the optical/infrared kilonova AT2017gfo \citep{Coulter2017,Chornock2017,Kasen2017,Pian2017,Smartt2017} and a broad-band afterglow \citep{Hallinan2017,Margutti2017,Troja2017,Mooley2018,Alexander2018},
established BNS mergers as one of the progenitors of short GRBs \citep{Eichler1989,Narayan1992,Berger2014} and major sites of $r$-process nucleosynthesis \citep{Lattimer1974,Metzger2010}. These observations provide a unique opportunity to study, within a single BNS event, the gravitational radiation, relativistic outflows, kilonova ejecta, and the physics of the post-merger remnant collectively. The event was localized to the nearby host galaxy NGC~4993,
with a spectroscopically measured heliocentric redshift of
$z=0.009783\pm0.000023$ \citep{Hjorth2017}.

Although the individual messengers from GW170817 have been extensively studied, a unified assessment of how the total energy of the event is distributed among its different physical channels remains lacking. A binary neutron star (BNS) merger partitions its mass--energy into several 
components, including gravitational wave emission \citep{Blanchet2014PNreview}, prompt and afterglow electromagnetic radiation \citep{Nakar2007,KumarZhang2015,Abbott2017MultiMessenger}, thermal and kinetic energy of the kilonova ejecta 
\citep{Metzger2017,Villar2017}, outflows powered by the accretion disk \citep{Perego2017,Siegel2018}, neutrino emission from the hot post-merger remnant and the outflows \citep{Dessart2009,Perego2014,2016PhRvD..94l3016F,IceCube2017}, and energy retained in the compact remnant through its gravitational mass and binding energy \citep{LattimerPrakash2001}.

In this work, we construct an Arnowitt--Deser--Misner (ADM) anchored, multi-channel energy budget for GW170817/GRB~170817A, adopting the ADM mass at the onset of the observable inspiral as a common energy reference \citep{Arnowitt1962}. We estimate the gravitational-wave radiated energy ($E_{\rm GW}$) from an independent Bayesian analysis of the GW170817 strain data, integrating the angle-resolved waveform energy over the full solid angle, and the prompt jet energy by 
accounting for its structure and viewing geometry. The afterglow and kilonova energetics, disk ($E_{\rm disk}$) and ejecta ($E_{\rm ej}$) masses are adopted from the reported literature. The total jet and kilonova constitutes the electromagnetic energy ($E_{\rm EM}$). The residual energy ($E_{\rm res}$) remaining after accounting for these channels encompasses neutrino losses ($E_{\nu}$) and energy 
retained by the compact remnant ($E_{\rm rem}$). This framework provides a global calorimetric closure of the event, allowing us to quantify how the available mass energy of the BNS merger is partitioned among its observable and unobservable channels.

The paper is organized as follows. Section~\ref{methodology} describes the methodology used to estimate the ADM reference energy and the energies associated with the gravitational wave, prompt jet, afterglow, and kilonova channels. Section~\ref{results} presents the resulting energy 
partition and the inferred neutrino energy and remnant mass. Section~\ref{discussion} discusses the implications of the global energy budget and its potential utility for constraining electromagnetic counterparts in future multimessenger BNS observations. Section~\ref{sec:summary} summarizes our main results.

\section{Methodology}
\label{methodology}

We adopt the effective ADM energy at the onset of the observable inspiral, $E_{\rm ADM}^{\rm eff}$, as the reference energy for the calorimetric closure. The energy budget is written as
\begin{equation}
E_{\rm ADM}^{\rm eff}
= E_{\rm GW} + E_{\rm EM} + E_{\rm disk}
+ E_{\rm ej} + E_{\rm res},
\end{equation}
where
\begin{equation}
E_{\rm res} \equiv E_{\nu}+E_{\rm rem}.
\end{equation}
The electromagnetic contribution comprises the intrinsic prompt jet emission, afterglow, and kilonova components,
\begin{equation}
E_{\rm EM}=E_{\gamma}+E_{\rm AG}+E_{\rm KN}.
\end{equation}
The afterglow ($E_{\rm AG}$) and kilonova ($E_{\rm KN}$) energetics of GRB170817A are taken from the detailed modelling of the electromagnetic counterparts available in literature, while the disk and ejecta contributions are based on the numerical relativity studies of the merger. Once these components are accounted for, the closure determines the residual budget $E_{\rm res}$. We then use baryon-mass conservation to constrain the remnant mass contribution, allowing the remaining residual energy to be attributed to neutrino emission. GW quantities come from Bayesian (nested-sampling) parameter estimation and are quoted as credible intervals; all other quantities are propagated by Monte Carlo with inputs drawn uniformly within adopted ranges and are quoted as central 68\% intervals.

The methodology and estimates adopted for each of these components are presented in detail in the following subsections.

\subsection{Estimate of effective ADM mass}
\label{adm_mass}
The total energy reservoir of the binary neutron-star merger is anchored to its
Arnowitt--Deser--Misner (ADM) mass \citep{Arnowitt1962}. Numerical 
relativity (NR) simulations typically begin at gravitational-wave frequencies
$f_{\rm NR}\gtrsim400$--$600$~Hz, above the onset of the observable GW170817
inspiral. However, we reconstruct the effective ADM mass at a reference frequency
$f_{\rm ref}=56.5$~Hz, corresponding to $\sim10$~s before the merger which represents the onset of the observable inspiral. We use the
publicly available CoRe/BAM NR database \citep{Dietrich2018CoRe}, selecting
configurations consistent with the GW170817 posterior and extracting their
initial ADM masses, component masses, tidal deformabilities, and starting GW
frequencies. The ADM mass is mapped from the NR starting frequency to
$f_{\rm ref}$ using the post-Newtonian binding energy; the detailed
construction is given in Appendix~\ref{app:adm_eff}. This yields
\begin{equation}
    E_{\rm ADM}^{\rm eff}(56.5~{\rm Hz})
    =
    (4.89\text{--}4.94)\times10^{54}\ {\rm erg},
    \label{eq:adm_eff_range}
\end{equation}
which we adopt as the initial energy reservoir for the global calorimetric
closure.

\subsection{Gravitational-wave energy}
\label{GW}

We use the publicly available GW170817 strain data from the LIGO
Hanford (H1) and Livingston (L1) detectors through the GWOSC archive
\citep{GWOSC2017}. Rather than propagating the public LIGO/Virgo
posterior samples, we perform an independent Bayesian parameter
estimation using \texttt{IMRPhenomPv2\_NRTidal}
\citep{PhysRevLett.113.151101, Dietrich2017NRTidal, Dietrich2019NRTidal}. We consider both low- and
high-spin priors (Table~\ref{tab:gw_pe_setup} in Appendix \ref{app:gw_energy}) and incorporate the
electromagnetic association with NGC~4993 through the sky position
and luminosity-distance priors, while retaining an isotropic prior
on the binary orientation. The resulting posterior samples provide
the binary parameters, including the component masses
($m_1,m_2$), dimensionless spin magnitudes ($a_1,a_2$), tidal
deformabilities ($\Lambda_1,\Lambda_2$), luminosity distance
($d_L$), and inclination ($\theta_{\rm JN}$), used in the
subsequent energy accounting.

For each posterior sample, we reconstruct the dominant quadrupolar
waveform modes using five independent viewing directions of the
same binary configuration. The total GW luminosity is then obtained
by integrating the corresponding mode power over the full solid
angle, accounting for the anisotropic emission of gravitational
waves. The source frame radiated energy is calculated by integrating
this angle integrated luminosity over the modeled waveform duration,
with the appropriate cosmological redshift correction. The
integration starts at $t_{\rm ref}=-10$~s and extends beyond the
waveform-amplitude peak until the cumulative modeled energy reaches
numerical saturation (reached $\approx10$~ms after the amplitude peak). This calculation includes the late-time
contribution of the generated waveform but does not independently
model the physical post-merger emission from the remnant. The
detailed methodology is presented in Appendix~\ref{app:gw_energy}. 

\begin{figure}[ht]
    \centering
    \includegraphics[width=1.0\linewidth]
    {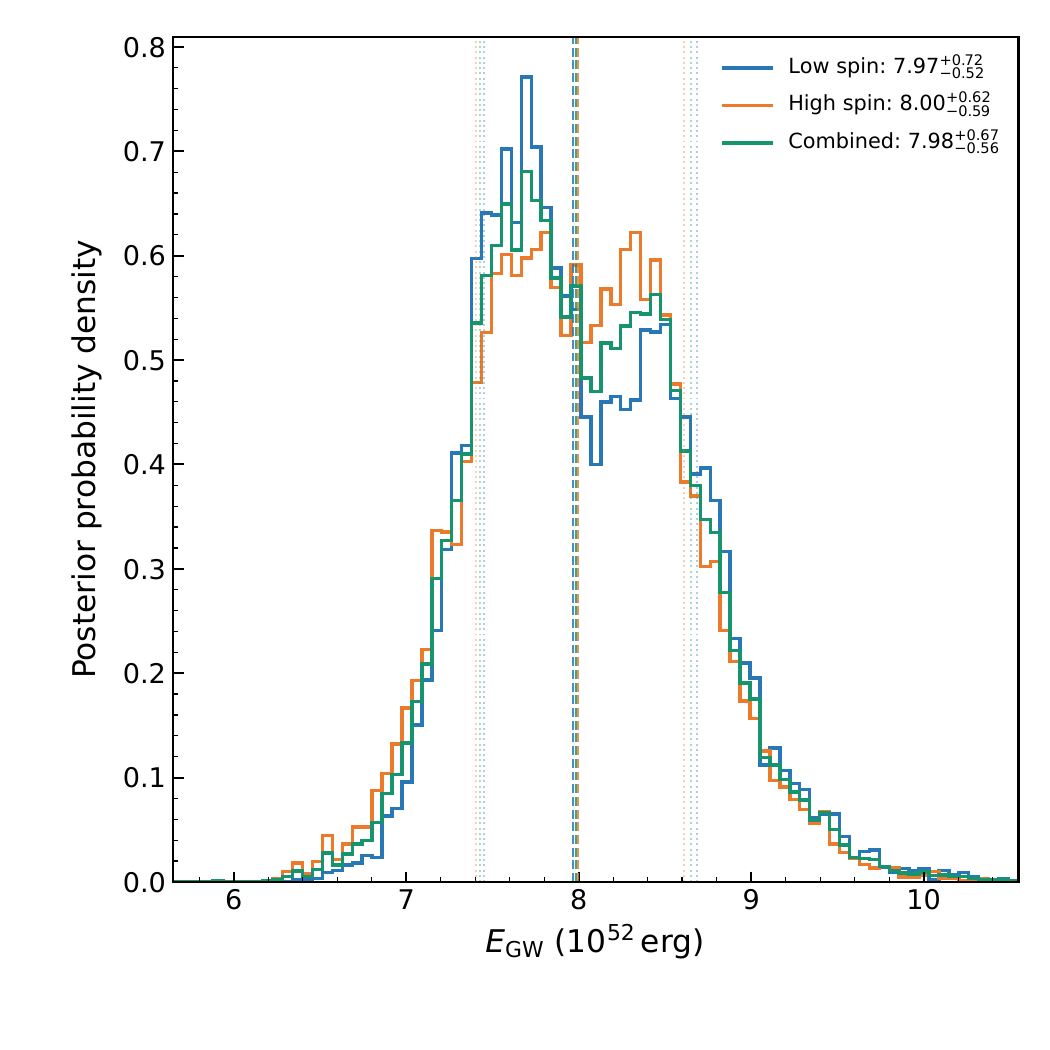}

    \caption{Posterior distribution of the angle-integrated, source-frame
    gravitational-wave energy radiated by GW170817 is shown.
    The low-spin (blue), high-spin (orange), and combined (green) distributions are shown, with vertical dashed and dotted lines marking the median and 68\% credible interval, respectively.}
    \label{fig:gw_energy_histogram}
\end{figure}

The low-spin ($a1, a2\equiv [0, 0.05]$) and high-spin ($a1, a2\equiv [0, 0.89]$) analyses yield consistent GW-energy
distributions. Combining the 9591 low-spin and 10518 high-spin
posterior samples, we obtain
\begin{equation}
E_{\rm GW}
=
7.98^{+0.67}_{-0.56}
\times10^{52}\ {\rm erg},
\label{eq:egw_final}
\end{equation}
where the uncertainties denote the 68\% credible interval of the
combined distribution. The corresponding posterior distribution is
shown in Figure~\ref{fig:gw_energy_histogram}. The resulting \(E_{\rm GW}\) posterior exhibits a bimodal structure, reflecting the non-Gaussian nature of 
the inferred energy distribution. We retain the full posterior when deriving the reported median and credible interval. Similar bimodality in the GW170817 tidal posterior has previously been discussed as potentially arising from the detector noise realization \citep{PhysRevX.9.011001}. We, therefore, do not interpret the two peaks as evidence for distinct physical
populations. We adopt this GW energy estimate in the subsequent ADM anchored calorimetric energy-budget analysis.

\subsection{Electromagnetic Counterparts (Prompt + Afterglow + Kilonova)}
\label{EM}
\begin{figure*}
    \centering
    \includegraphics[width=7.0in]{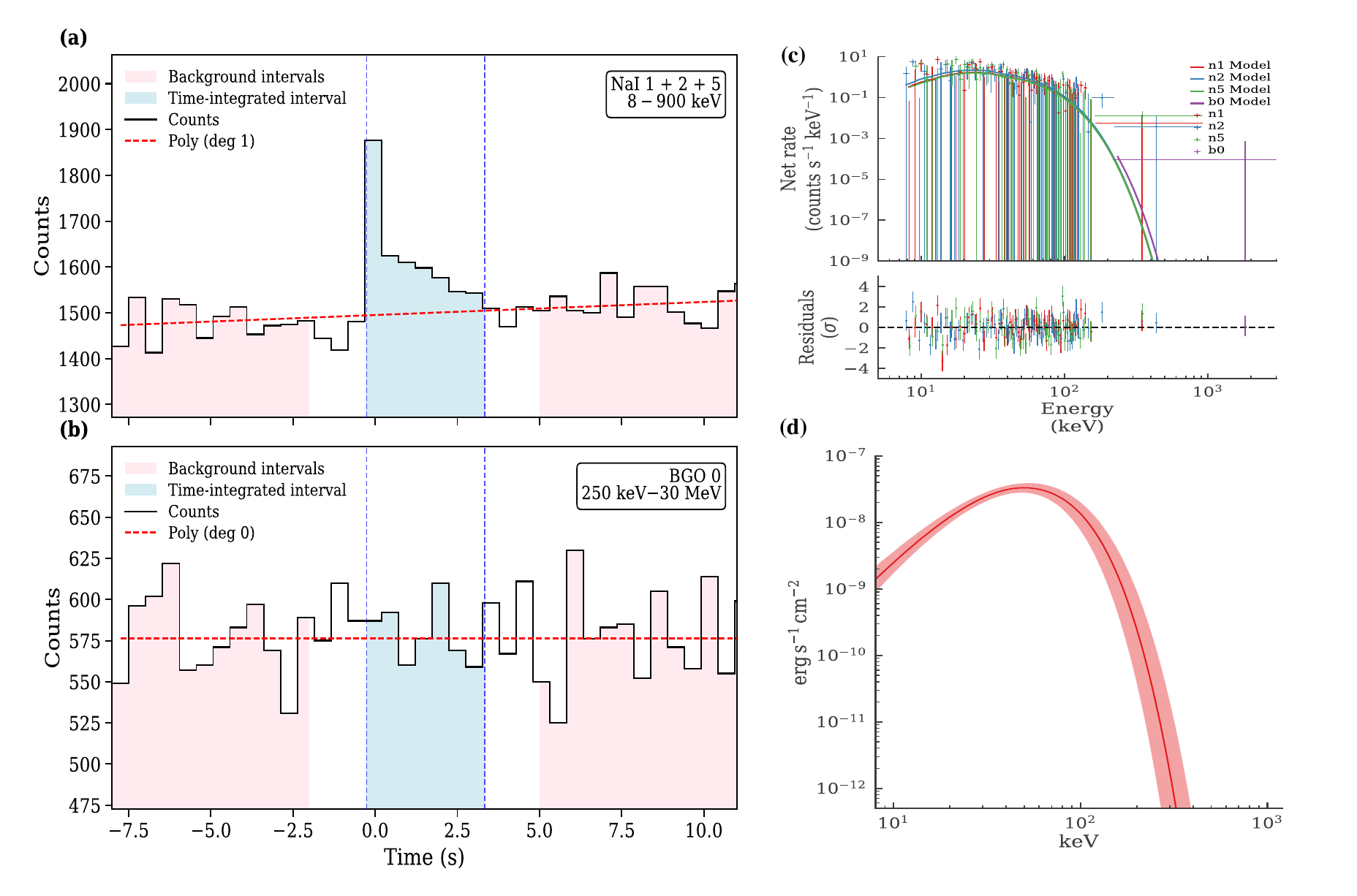}

    \caption{GBM count light curves of GRB~170817A in (a) NaI~1+2+5
    (8--900~keV) and (b) BGO~0 (250~keV--30~MeV). The blue shaded region
    denotes the source interval, $t \in [-0.258, +3.328]$~s, while the pink
    regions indicate the background intervals, $[-8, -2]$~s and
    $[+5, +11]$~s. The horizontal dashed curves represent the background polynomial fits.
    (c) Jointly fitted time-integrated count spectra and residuals for the
    cutoff power-law model. (d) Corresponding $\nu F_\nu$ spectrum, with the
    shaded region indicating the propagated 68\% uncertainty in the
    estimated model shape.}    
    \label{fig:grb170817a_spectrum}
\end{figure*}

\noindent
{\bf Prompt Emission}\\
GRB~170817A was detected by the \textit{Fermi} Gamma-ray Burst Monitor (GBM)
$1.734$~s after the GW merger time of GW170817 \citep{Goldstein2017Fermi, Abbott2017GRB}.
We use the GBM time-tagged event (TTE) data from NaI~1, NaI~2, and NaI~5,
covering $\sim8$--$900$~keV, together with BGO~0, covering
$\sim250$~keV--$30$~MeV \citep{Meegan2009}. The prompt emission spectrum is analyzed using the Maximum Likelihood
Estimation (MLE) method within 3ML \citep{Vianello2015}, over
$t=-0.258$ to $+3.328$~s relative to the GBM trigger, encompassing the burst duration (Figures~\ref{fig:grb170817a_spectrum}a and \ref{fig:grb170817a_spectrum}b). The time-integrated spectrum over this interval is described by a
cutoff power-law model, yielding a 10--1000~keV fluence of $
F_{\rm obs}=(1.80^{+0.28}_{-0.27})\times10^{-7}\ {\rm erg\,cm^{-2}}$ (Appendix \ref{app:prompt_spectral}). Figures~\ref{fig:grb170817a_spectrum}c and \ref{fig:grb170817a_spectrum}d show the count spectra with residuals and the $\nu F_\nu$ spectrum of the cutoff power-law model. 
Using the GW inferred distance to NGC~4993, this corresponds to an observed
isotropic-equivalent prompt energy $
E_{\rm iso,obs}
=(3.48^{+0.90}_{-0.53})\times10^{46}\ {\rm erg}.
$

The burst was observed substantially off-axis, with an adopted viewing-angle range of $\theta_{\rm obs}=15^\circ$--$25^\circ$, consistent with constraints from the afterglow and VLBI observations \citep{Mooley2018, Ghirlanda2019}. We, therefore, apply 
structured-jet corrections to the observed isotropic-equivalent energy to estimate the true prompt emission. We account for this geometric effect using 
Gaussian and power-law structured-jet models (power law index, $k=4-6$), with the jet-core scale constrained through $\theta_{\rm obs}/\theta_c=3.5$--$6.5$. The resulting jet-integrated prompt radiated energies are
\begin{equation}
\begin{array}{l}
E_{\gamma}^{\rm (Gauss)}
=3.78^{+56.30}_{-3.44}\times10^{49}\ {\rm erg},\\[8pt]
E_{\gamma}^{\rm (PL)}
=1.84^{+3.05}_{-1.08}\times10^{47}\ {\rm erg}.
\end{array}
\end{equation}
where the quoted values denote the median and central 68\% interval from Monte Carlo sampling uniformly over the adopted parameter ranges.
We, therefore, adopt
\begin{equation}
E_{\gamma}
=(7.6\times10^{46}-6\times10^{50})\ {\rm erg}
\end{equation}
as the prompt emission contribution to the global energy budget, with the range capturing the uncertainty associated with the assumed jet angular structure. The detailed calculations are given in Appendix \ref{app_jet_struc}\\
\noindent
{\bf Afterglow and Kilonova}\\
The broadband afterglow of GRB~170817A is consistent with a structured
relativistic jet. Multi-wavelength modelling gives a beaming corrected
blast-wave kinetic energy of the GRB jet to be
$E_{\rm AG}\sim 10^{50} - 10^{52}$ ~erg
\citep{Mooley2018, 10.1093/mnras/staa2626}. We adopt this published outflow energy as the afterglow contribution, without performing an independent afterglow analysis. We note that the prompt and afterglow emissions allow us to estimate the total energy available in the GRB jet.  

The kilonova AT2017gfo associated with GW170817 has a total radiated
bolometric energy of approximately
$E_{\rm KN}\simeq10^{47}$~erg over the first $\sim10$ days \citep{Cowperthwaite_2017,Villar2017,Kasen2017}. We adopt this value as the kilonova contribution to the electromagnetic energy budget.


\subsection{Disk and Ejecta Mass}
\label{disk_ejecta}

The baryonic masses of the dynamical ejecta and post-merger accretion disk
are adopted as inputs to the mass and global energy budgets. For the ejecta, we adopt the range
$M_{b,\rm ej} \in [\,0.056,\ 0.064\,]\,M_\odot,$
consistent with multi-component kilonova modelling of AT2017gfo
\citep{Cowperthwaite_2017, Villar2017, Perego2017}. 

The disk mass is not directly constrained by
electromagnetic observations and is, therefore, taken from numerical-relativity
studies of binary neutron-star mergers with long-lived remnants
\citep{Radice_2018,Nedora2021} $M_{b,\rm disk} \in [\,0.10,\ 
0.20\,]\,M_\odot$. The adopted ejecta and disk masses are baryonic masses reported in the
literature. We, thereby, formulate their contribution to the global bookkeeping in terms of baryonic rest-mass energy, $E_{\rm rest}=Mc^2$, neglecting separate kinetic, thermal, and binding-energy corrections at leading order \citep{Salafia2022}. A conversion to gravitational mass is not uniquely
known for the adopted mass estimates, as it depends on the equation-of-state (EOS) and the physical configuration of the ejecta and disk. Thus, the corresponding energies are considered to be: $E_{\rm ej}\simeq (1.00-1.14)\times10^{53}\ {\rm erg} $ and $E_{\rm disk} \simeq (1.79-3.57)\times10^{53}\ {\rm erg}$. 



\section{Results}
\label{results}
\subsection{Constraints on the Remnant and Neutrino Emission}
\label{sec:remnant_neutrino}

Having accounted for the gravitational wave, electromagnetic, disk, and
ejecta contributions, we define the remaining energy as the residual energy,
\begin{equation}
\label{eq:Eres_def}
E_{\rm res}
=
E_{\rm ADM}^{\rm eff}
-
E_{\rm GW}
-
E_{\rm EM}
-
E_{\rm disk}
-
E_{\rm ej}.
\end{equation}
Using the energy ranges estimated in section \ref{methodology}, we obtain
\begin{equation}
\label{eq:Eres_range}
E_{\rm res}
\in
[\,4.32,\ 4.59\,]\times10^{54}\ {\rm erg}.
\end{equation}
This residual energy is what is available to be partitioned between the gravitational mass-energy retained in the compact remnant and the energy carried away by neutrinos.

\subsubsection{Baryonic-mass bookkeeping and remnant mass}
\label{sec:remnant_mass_constraint}
\begin{figure*}
    \centering
    \includegraphics[width=7.0in]{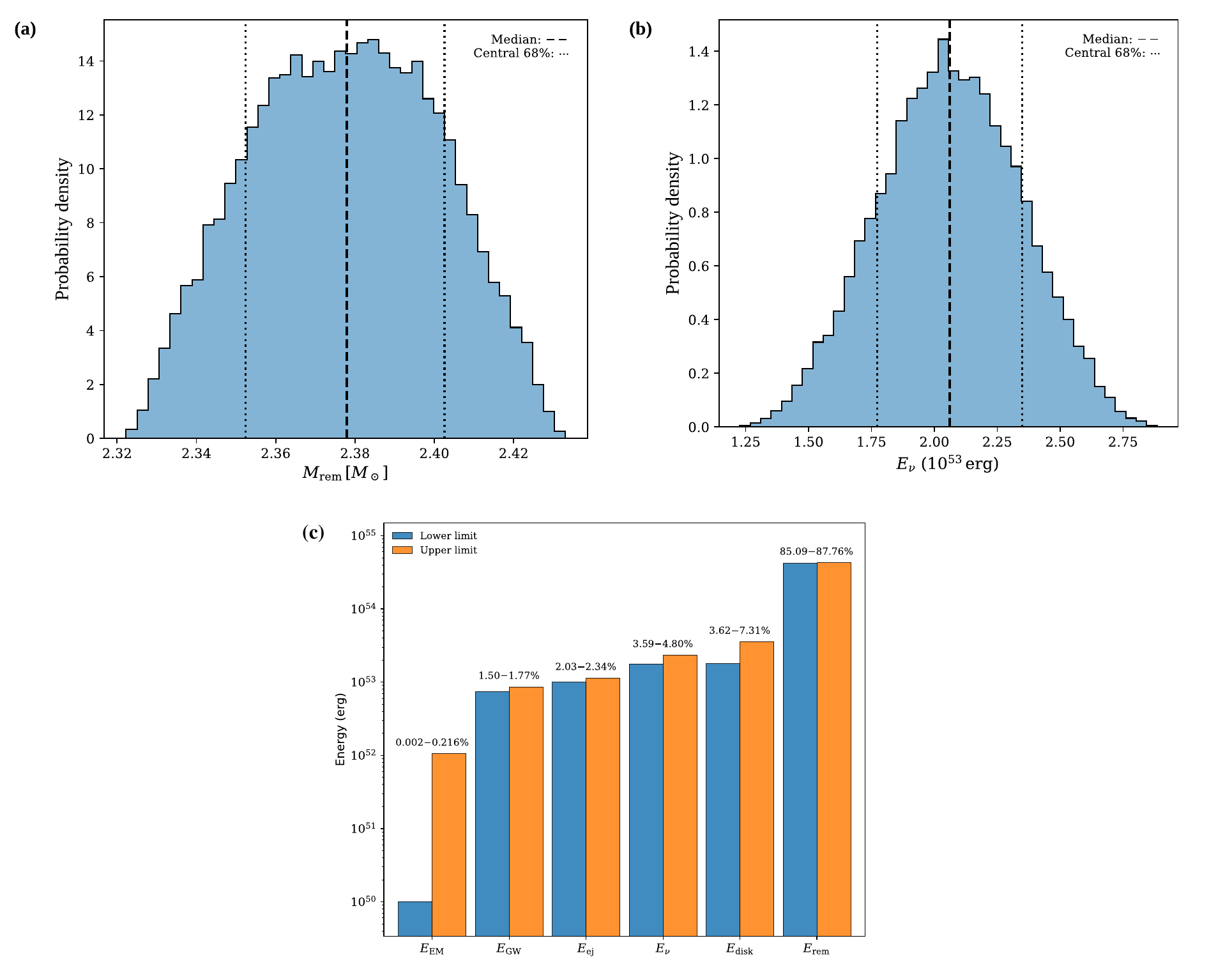}
    \caption{
       Distributions of (a) the compact-remnant gravitational mass, with a median
of $M_{\rm rem}=2.378^{+0.025}_{-0.026}\,M_\odot$, and (b) the total
neutrino energy inferred from the calorimetric energy balance, with a
median of $E_\nu=2.061^{+0.287}_{-0.288}\times10^{53}$~erg, are shown. (c)  Channel-wise energy partition of GW170817, normalized to the ADM-effective energy reservoir at $t_{\rm ref}=-10$~s. Blue and orange bars denote the lower and upper estimates, respectively. 
    Each channel's bounds are estimated independently and do not represent a joint energy-budget realization.
    }
    \label{fig:remnant_mass_neutrino}
\end{figure*}

Following the merger, we consider that the bound compact core immediately remains a rapidly rotating neutron-star remnant. Depending on the maximum mass supported by the equation 
of state (EOS) and the degree of rotational support, the remnant may be either supramassive or hypermassive and may subsequently
collapse to a black hole. Our energy bookkeeping refers to this immediate post-merger compact remnant and does not require specifying its eventual fate. We, therefore, estimate its gravitational mass from its conserved baryonic mass using the EOS insensitive baryonic-gravitational mass relation for rotating neutron stars \citep{2020FrPhy..1524603G} (Appendix~\ref{app:baryonic_bookkeeping}). The same relation is applied to the progenitor stars, using the combined low- and high-spin GW posterior samples, and to the remnant.

We first constrain the remnant independently of the energy partition by conserving the baryonic mass. For each gravitational wave posterior sample, the baryonic masses of the two progenitor stars are obtained from this relation. The baryonic mass of
the bound post-merger remnant is then
\begin{equation}
\label{eq:Mb_rem_main}
M_{b,\rm rem}
=
M_{b,1}+M_{b,2}
-
M_{b,\rm ej}
-
M_{b,\rm disk}.
\end{equation}
The resulting remnant baryonic mass is converted to gravitational mass by inverting the same relation, which gives
\begin{equation}
\label{eq:Mrem_main}
M_{\rm rem}
\in
[\,2.352,\ 2.403\,]\,M_\odot
\qquad (\rm \ 68\% \ central \ interval).
\end{equation}
The corresponding distribution of the compact remnant gravitational mass is shown in Figure \ref{fig:remnant_mass_neutrino}a.
The corresponding gravitational mass-energy retained in the compact remnant
is, therefore,
\begin{equation}
\label{eq:Erem_main}
E_{\rm rem}
=
M_{\rm rem}c^2
\in
[\,4.20,\ 4.29\,]\times10^{54}\ {\rm erg}.
\end{equation}
The detailed central 68\% interval is presented in Appendix~\ref{app:baryonic_bookkeeping}.

\subsubsection{Constraint on neutrino emission}
\label{sec:neutrino_constraint}

The residual energy not retained as gravitational mass-energy of the compact remnant provides a calorimetric constraint on the total neutrino emission,
\begin{equation}
\label{eq:Enu_main}
E_\nu
=
E_{\rm res}
-
M_{\rm rem}c^2.
\end{equation}
Propagating the residual energy and remnant mass distributions gives
\begin{equation}
\label{eq:Enu_main_range}
E_\nu
\in
[\,1.77,\ 2.35\,]\times10^{53}\ {\rm erg}
\qquad (68\% \, \rm central \, interval).
\end{equation}
The corresponding distribution of the calorimetrically inferred total neutrino energy $E_\nu$ is shown in Figure \ref{fig:remnant_mass_neutrino}b. 


\subsection{Global Energy Partition of GW170817}
\label{sec:global_energy_partition}

Using the ADM-effective energy at $t_{\rm ref}=-10$~s as the reference energy reservoir, we obtain the global energy partition of GW170817 across the compact remnant, gravitational waves, 
disk, ejecta, neutrinos, and electromagnetic radiation. The compact remnant dominates the energy budget, retaining
$\simeq85.09$--$87.76\%$ of the ADM-effective energy.
The baryonic disk and ejecta account for
$\simeq3.62$--$7.31\%$ and $\simeq2.03$--$2.34\%$,
respectively, while the neutrino channel contributes
$\simeq3.59$--$4.80\%$. Gravitational wave emission
carries $\simeq1.50$--$1.77\%$, whereas the
electromagnetic associated energy budget contributes only
$\simeq0.002$--$0.216\%$
(Figure~\ref{fig:remnant_mass_neutrino}c).
Individually, the jet-integrated prompt $\gamma$-ray emission
accounts for $\simeq1.5\times10^{-6}$--$0.0123\%$,
the afterglow blast-wave kinetic energy for
$\simeq0.002$--$0.204\%$, and the kilonova bolometric
radiation for $\sim2.0\times10^{-6}\%$ of the ADM-effective energy.


The resulting hierarchy shows that the global merger energy is overwhelmingly retained in the compact remnant, with the disk, ejecta, neutrinos, and gravitational waves constituting the 
principal subdominant channels. In contrast, the observed electromagnetic emission represents only a negligible fraction of the total energy reservoir. The quoted ranges represent
the independently allowed limits and, therefore, are not required to sum to unity at their extrema.

\section{Discussion \& Implications}
\label{discussion}
\begin{figure*}
    \centering
    \includegraphics[width=7.0in]{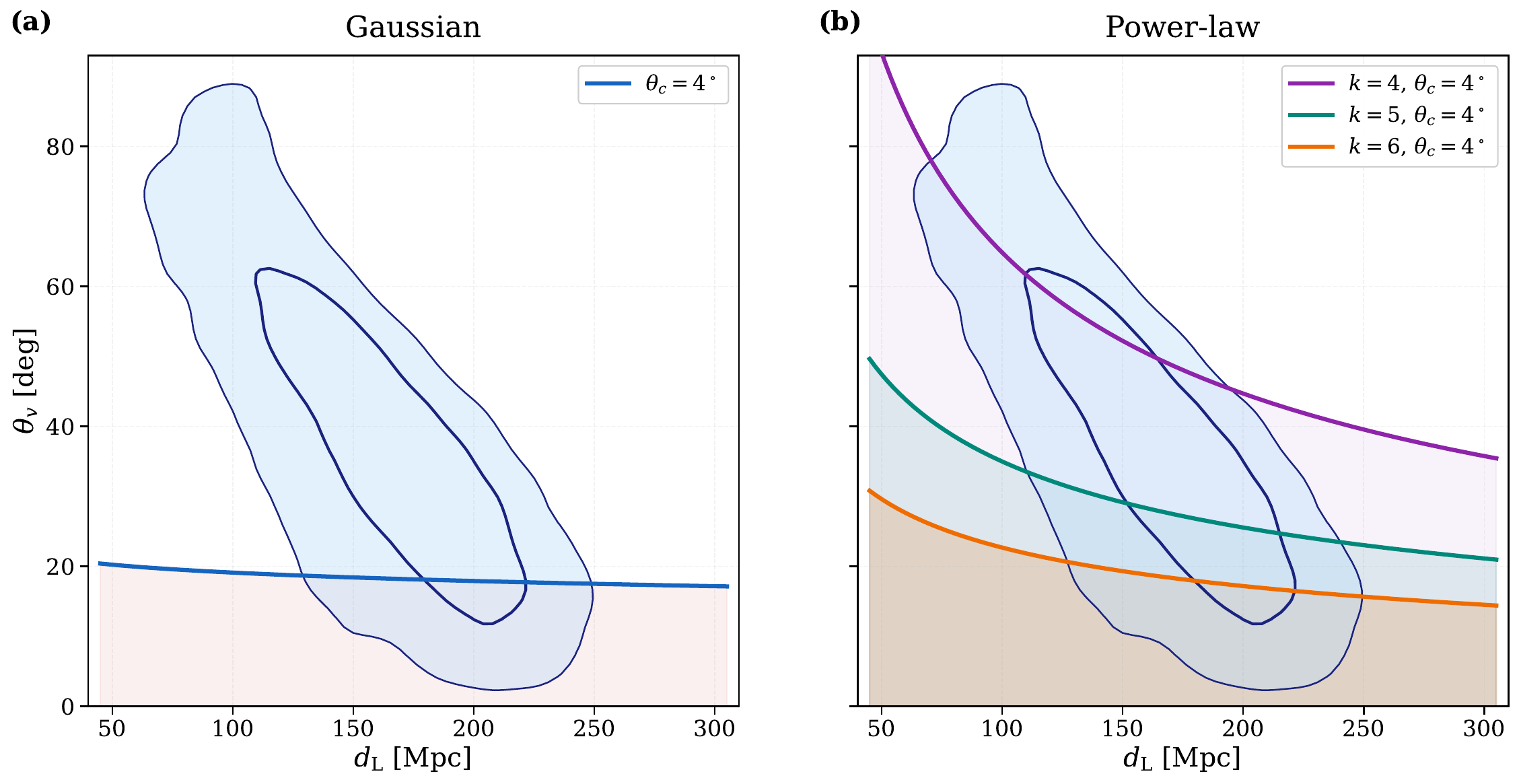}
    \caption{
    Constraints imposed on the GW-derived $d_L$--$\theta_v$ parameter space of GW190425 by the \textit{Fermi}-GBM non-detection, using the GW170817 calibrated prompt energy fraction for (a) Gaussian and (b) power-law jet structures are shown.
    The solid curves show the $\theta_{v,\min}(d_L)$ required for the prompt fluence to fall below the GBM detection threshold at each luminosity distance. The shaded regions below the curves correspond to viewing angles for which the prompt emission would have been detectable and are, therefore, disfavoured by the non-detection.
    }
    \label{fig:gw190425_gbm_constraint}
\end{figure*}

\subsection{Implications of the IceCube Non-detection}
\label{sec:icecube_discussion}

Neutrino emission associated with a binary neutron-star merger can arise through physically distinct channels. A substantial neutrino flux is expected from the hot, dense remnant and surrounding accretion disk formed immediately following the merger. In these regions, neutrinos are predominantly produced
through thermal processes and are expected to have characteristic energies ($\sim 10-25 \, \rm MeV$) much lower than those of the high-energy neutrinos targeted by IceCube \citep{Rosswog_etal_2003,Radice_etal_2020}. A second channel is associated with the relativistic outflow responsible for the prompt GRB emission \citep{IceCube2017}. In this case, protons accelerated in the dissipation
region can interact with the ambient photon field through photohadronic processes, producing pions and subsequently high-energy, non-thermal neutrinos. Such prompt GRB neutrinos are among the high-energy neutrino components targeted in searches by IceCube \citep{Icecube_GRB_searches_2022}.

Our energy-partition analysis constrains the total neutrino energy to be of order $10^{53}$~erg. This constraint is obtained from the global energy balance and, therefore, includes the neutrino energy irrespective of its production channel or characteristic energy. In contrast, the non-detection of high-energy neutrinos associated with GW170817 by IceCube
and other neutrino observatories places an upper limit on the
high-energy, non-thermal component. For the assumed $E^{-2}$ neutrino spectrum, the IceCube non-detection
constrains the all-flavor high-energy neutrino energy to
$\sim(1.6 - 2.0)\times10^{51}$~erg for the adopted distance \citep{IceCube2017}.

The high-energy neutrino constraint is substantially below the total neutrino energy inferred from our energy budget. Thus, the non-detection 
implies that only a small fraction of the inferred neutrino energy could have been carried by the high-energy non-thermal component associated with the relativistic GRB
outflow. The bulk of the neutrino energy inferred from our global energy balance is, therefore, more naturally associated with the thermal neutrino emission from the hot post-merger remnant and accretion disk, which is not directly constrained by the IceCube high-energy non-detection. However, the non-detection of MeV neutrinos by Super-Kamiokande is also consistent with our inferred neutrino energy budget, given the large distance to GW170817 ($\sim40$ Mpc), which results in a neutrino fluence well below its detection sensitivity \citep{Abe_etal_2018}.

\subsection{Implications of Post-merger Gravitational-wave Emission}
\label{sec:postmerger_GW}

Our global energy budget does not explicitly account for gravitational wave
emission from the post-merger remnant. We, therefore, assess whether this
omission could significantly affect the inferred neutrino energy budget.
Using the remnant mass constrained in Section~\ref{sec:remnant_neutrino},
we estimate the rotational energy reservoir of the newly formed neutron
star. The available rotational energy is
\begin{equation}
E_{\rm rot}=\frac{1}{2}I\Omega^2
=\frac{2\pi^2 I}{P^2},
\end{equation}
where $I$ and $P$ are the moment of inertia and rotation period,
respectively. We adopt $M_{\rm rem}=2.352$--$2.403\,M_\odot$ and
a representative radius range of $R=11$--$13$~km. Using the approximate
moment of inertia relation $I\simeq0.35M_{\rm rem}R^2$
\citep{LattimerPrakash2001}, 
and rotation periods of $0.8$ - $1$~ms \citep{Chakrabarty_2008,Papitto_etal_2014}, representative of a rapidly
rotating newly born neutron star, the corresponding rotational-energy
reservoir is $E_{\rm rot}\simeq(3.9$--$8.7)\times10^{52}$~erg.

Only a fraction of this rotational energy is expected to be carried away
by gravitational waves. Thus, the actual post-merger GW contribution is
expected to be smaller than the estimated rotational energy reservoir.
Furthermore, for a supramassive or hypermassive, remnant, the eventual collapse to a black
hole terminates the neutron star spin-down phase, while the subsequent
black hole ringdown is expected to contribute negligibly to the energy
budget. The main additional GW losses would, therefore, arise during the
preceding neutron star phase, through non-axisymmetric instabilities \citep{Metzger2017}.
We, thus, expect the omission of post-merger GW emission to have a limited
impact on our inferred neutrino energy budget of
$E_\nu\sim10^{53}$ erg.

\subsection{Application to GW190425: constraining the hidden prompt counterpart}
\label{subsec:gw190425_gbm_veto}

The energy partition inferred from GW170817 can provide an electromagnetic constraint on the distance-inclination parameter space of GW only events. As a demonstration, we consider GW190425 \citep{Abbott2020GW190425, Fletcher2019GBM}, for which the GW signal alone leaves a strong 
degeneracy between luminosity distance and inclination. We ask how the absence of a \textit{Fermi}-GBM counterpart constrains this GW-derived, luminosity distance - viewing angle ($ d_L$--$\theta_v$) posterior if the jet energy fraction is comparable to that inferred from GW170817.

We adopt the GW170817 calibrated average jet energy fraction, $f_{\rm jet}=0.11\%$, which accounts for both prompt and afterglow emission (Section~\ref{sec:global_energy_partition}), and assume an average prompt radiative efficiency of $0.15$ for short GRBs \citep{Beniamini2016, OConnor2024}. This yields a reference prompt-energy fraction of $f_{\gamma}=0.016\%$. Applying this fraction to $M_{\rm ADM}\simeq 3.4\,M_\odot$ for GW190425 gives an on-axis isotropic equivalent prompt energy of $E_0\simeq 9.7\times10^{50}$ erg.
This corresponds to an on-axis GBM detection horizon of $d_{\rm crit,on}\simeq2.96\times10^{4}$ Mpc, vastly exceeding the GW190425 distance range, $d_L=159^{+69}_{-71}$ Mpc constrained by the GW analysis contour (Figure~\ref{fig:gw190425_gbm_constraint}). Hence, a near-axis jet with this
energy scale cannot be hidden by distance; the GBM non-detection, thereby, excludes the portion of the GW posterior corresponding to sufficiently small inclination angles.

For each distance, we determine the minimum inclination $\theta_{v,\min}(d_L)$ required 
for the prompt fluence to fall below the GBM threshold, shown in solid line in Figure~\ref{fig:gw190425_gbm_constraint}. For the Gaussian profile, the GBM non-detection rules out viewing angles $\theta_v\lesssim18^\circ$ across the entire luminosity-distance range spanned by the GW posterior. For power-law profiles with $\theta_c=4^\circ$, the inferred lower limits on $\theta_v$ depend strongly on the wing slope: $\theta_v\gtrsim40^\circ$--$80^\circ$ for $k=4$, $23^\circ$--$37^\circ$ for $k=5$, and $16^\circ$--$24^\circ$ for $k=6$, with the limits decreasing with increasing $d_L$.

Thus, the energy partition provides an independent electromagnetic constraint on the GW inferred distance-inclination degeneracy, excluding the portion of the GW posterior corresponding to a sufficiently near-axis successful jet. Here, however, the jet energy fraction is calibrated from a single event, 
GW170817. We assess the sensitivity to the adopted jet-energy fraction by varying it by a factor of two, finding that the Gaussian constraint remains relatively insensitive to this variation, whereas the shallower power-law profiles exhibit substantially greater sensitivity (Appendix~\ref{app:gw190425_fgamma_sensitivity}). With a larger sample of mergers with identified counterparts, the corresponding energy partitions can be established across events, allowing the merger energy partition, and hence, such electromagnetic constraints, to be tested and eventually calibrated empirically rather than inferred from a single system.

\section{Summary}
\label{sec:summary}

We have presented the first calorimetric energy partition of the binary neutron-star merger GW170817 associated with GRB 170817A, placing the gravitational-wave, prompt-emission, ejecta, and remnant energy budgets on a common scale relative to the pre-merger ADM energy. We find that gravitational-wave emission accounts for $\sim 1.6\%$ of the total energy, the prompt $\gamma$-ray emission contributes only $\lesssim 0.012\%$, while the disk and dynamical ejecta together contain $\sim 5$--$10\%$. The dominant fraction of the energy, therefore, remains in the compact-remnant and neutrino reservoirs.

This calorimetric budget provides direct constraints on the post-merger outcome. The inferred compact remnant mass is $2.352$--$2.403\,M_\odot$ 
and the neutrino energy is $(1.77$ - $2.35)\times10^{53}\,\mathrm{erg}$. The energy partition, thus, links the observed prompt emission and ejecta to the otherwise inaccessible energy reservoirs of the post-merger system, providing quantitative constraints on the remnant and its neutrino emission.

Importantly, the calorimetric constraints also narrow the luminosity distance-inclination parameter space permitted by the gravitational wave signal alone. The energetics of the merger, therefore, provide information complementary to the GW waveform, demonstrating that energy partitioning can help constrain degeneracies in 
the inference of binary geometry. GW170817 thus provides the first demonstration of calorimetric energy partitioning as a multimessenger diagnostic of neutron-star mergers, connecting the merger energy reservoir, remnant outcome, relativistic outflow, and prompt electromagnetic emission. Applying this framework to future events will enable these partitions to be tested across a population and provide new constraints on the physics of merger remnants and their central engines.

\begin{acknowledgments}
S.I. is supported by DST INSPIRE Faculty Scheme (IFA19-PH245) and ANRF ARG Grant (ANRF/ARG/2025/000380/PS). 
We acknowledge the support of High Performance Computing Centre (CHPC) of IISER TVM for providing the computational resources. 
\end{acknowledgments}

\begin{contribution}

R.S. carried out the formal analysis, developed the methodology, curated the data, generated the visualizations, and prepared the original draft of the manuscript. S.I. contributed to the conceptualization, investigation, and methodology, and was responsible for project administration and funding acquisition. She also contributed to writing, reviewing, and editing the manuscript. S.V.A. contributed to the formal analysis and review of the manuscript.


\end{contribution}

%
\facilities{Fermi (GBM), LIGO}

\software{Bilby \citep{bilby_paper}, parallel\_bilby \citep{pbilby_paper},
          dynesty \citep{dynesty_paper}, LALSuite \citep{lalsuite},
          3ML \citep{Vianello2015}}

ChatGPT (OpenAI GPT-6) was used to assist with language refinement, proofreading and improving presentation of selected portions of the manuscript. All scientific content, calculations, data analysis, interpretations and conclusions were independently carried out by the authors who take full responsibility for the manuscript. 

\appendix

\section{Construction of the Effective ADM Mass}
\label{app:adm_eff}

We construct the effective ADM energy at the reference gravitational wave frequency $f_{\rm ref}=56.5$~Hz using numerical-relativity (NR) configurations from the publicly available CoRe/BAM database \citep{Dietrich2018CoRe}. We retain configurations consistent with the GW170817 posterior (section \ref{GW}) and use the archived initial data quantities for each configuration: the ADM mass $M_{\rm ADM}(f_{\rm NR})$, component masses $(m_1,m_2)$, tidal deformabilities $(\Lambda_1,\Lambda_2)$, and initial GW frequency $f_{\rm NR}$. The BAM simulations typically begin at $f_{\rm NR}\gtrsim400$--$600$~Hz and therefore do not directly provide the ADM energy at the onset of the observable inspiral.

We map the ADM mass from $f_{\rm NR}$ to $f_{\rm ref}$ using the gauge invariant post-Newtonian (PN) binding energy,
\begin{equation}
    E_{\rm bind}(f)
    =
    E_{\rm pm}^{\rm 3PN}(f)
    +
    E_{\rm tidal}(f),
    \label{eq:app_bind}
\end{equation}
where $E_{\rm pm}^{\rm 3PN}$ is the point-mass contribution through 3PN order and $E_{\rm tidal}$ is the leading-order tidal contribution \citep{Flanagan2008,Vines2011}. For each NR configuration, we calculate
\begin{equation}
    \Delta E_{\rm bind}
    =
    E_{\rm bind}(f_{\rm ref})
    -
    E_{\rm bind}(f_{\rm NR}),
\end{equation}
and define
\begin{equation}
    M_{\rm ADM}^{\rm eff}(f_{\rm ref})
    =
    M_{\rm ADM}(f_{\rm NR})
    +
    \frac{\Delta E_{\rm bind}}{c^2}.
    \label{eq:app_admeff}
\end{equation}
Here, the GW frequency is related to the orbital frequency by $\Omega=\pi f$ for the dominant quadrupolar mode. This procedure accounts for the change in binary binding energy between the NR starting frequency and the reference frequency without extrapolating the NR simulations to lower frequencies.

We assess the systematic uncertainty associated with this PN correction using
the 18 CoRe hybrid waveforms considered in our analysis. For each configuration,
we compare the GW energy radiated over the same frequency interval,
$56.5~{\rm Hz}\rightarrow f_{\rm NR}$, as obtained from the 3PN+tides
prescription and from direct integration of the energy flux of the hybrid
$h_{22}$ waveform \citep{Flanagan2008, Vines2011, Wade2014, PhysRevLett.113.151101, Kiuchi2017}. The PN estimate exceeds the corresponding
hybrid-NR energy by a factor of $\simeq1.7$--$2.0$, with a median ratio of
approximately $2$. The resulting difference in the reconstructed energy is typically $\sim(1$--$2)\times10^{52}$~erg, which is $\lesssim0.4\%$ of $E^{\rm eff}_{\rm ADM}$ but comparable to the statistical width of $E_\nu$. Replacing the PN correction with the hybrid-NR calibration would lower $E^{\rm eff}_{\rm ADM}$, and hence affect the estimate of $E_\nu$, by this amount; we, therefore, treat it as a one-sided systematic on $E_\nu$ (Section~\ref{sec:neutrino_constraint}).

Converting the resulting masses to energy through
$E_{\rm ADM}^{\rm eff}=M_{\rm ADM}^{\rm eff}c^2$, we obtain
\begin{equation}
    E_{\rm ADM}^{\rm eff}(56.5~{\rm Hz})
    =
    (4.89\text{--}4.94)\times10^{54}\ {\rm erg},
    \label{eq:app_adm_range}
\end{equation}
where the range reflects the variation among the selected NR configurations.
We adopt this interval as the initial ADM-energy reservoir for the calorimetric
closure.

\section{Details of the GW-energy calculation}
\label{app:gw_energy}

\begin{figure*}[t]
    \centering
    \includegraphics[width=1.0\textwidth]{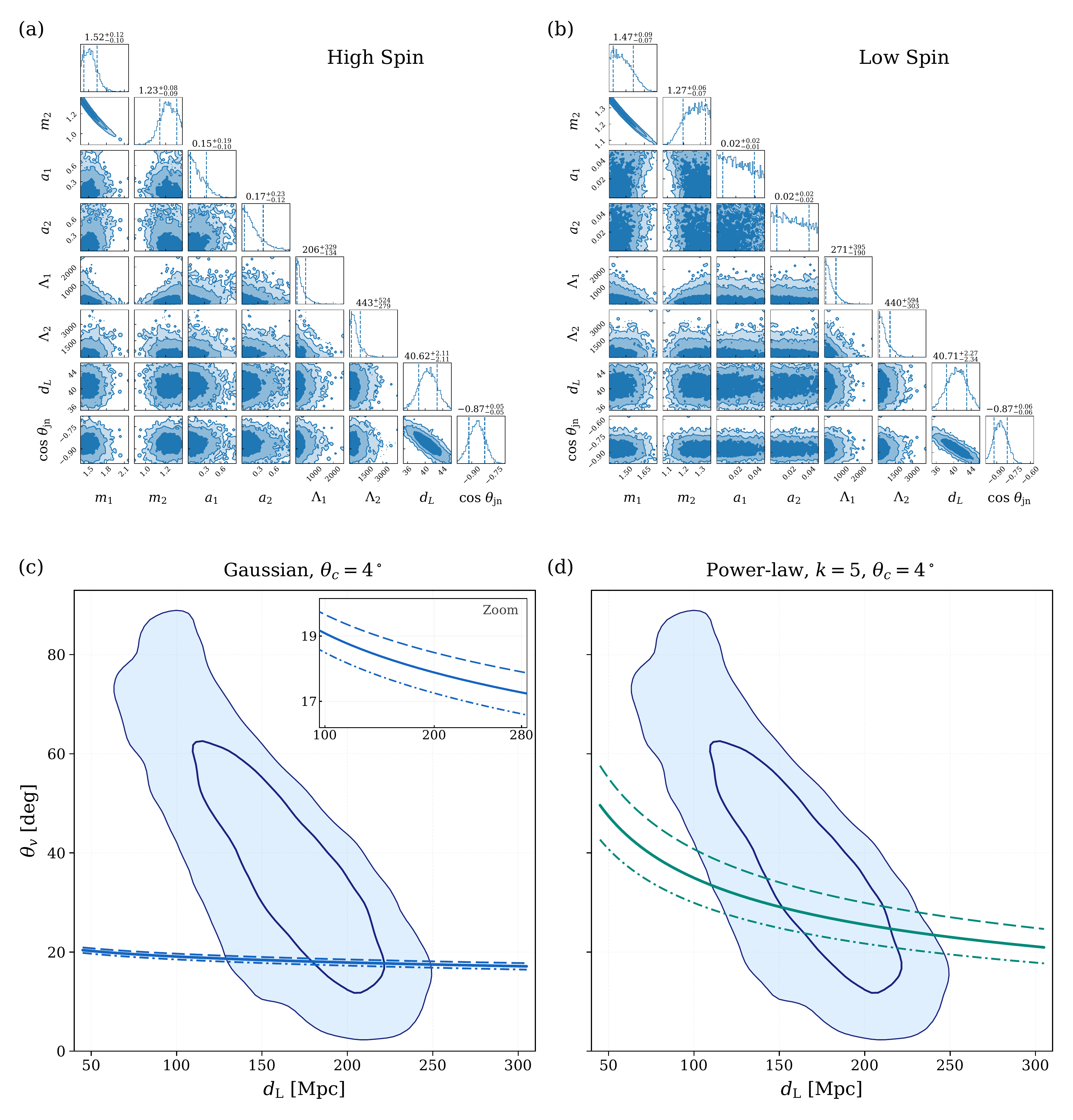}
    \caption{
        The posterior distributions of the binary parameters obtained
        in the GW analysis for the high-spin (a) and low-spin (b) cases
        are shown. The parameters shown are the component masses
        $(m_1,m_2)$, dimensionless spin magnitudes $(a_1,a_2)$, tidal
        deformabilities $(\Lambda_1,\Lambda_2)$, luminosity distance
        $d_L$, and inclination parameter $\cos\theta_{\rm JN}$. The values above the diagonal panels give the median and the 68\% credible interval of each parameter. Sensitivity of the conditional GW190425 viewing angle constraints to a factor of two variation in the jet energy fraction for a (c) Gaussian jet and a (d) power-law jet with $k=5$, both with $\theta_c=4^\circ$ are shown. Solid, dash-dotted and dashed curves correspond to $f_{jet}$, $0.5\,f_{jet}$ and $2\,f_{jet}$, respectively. Contours and shading follow Figure~\ref{fig:gw190425_gbm_constraint}.    }
    \label{fig:gw_posteriors_high_low_spin}
\end{figure*}

For each posterior sample (Figure~\ref{fig:gw_posteriors_high_low_spin})
described in Section~\ref{GW}, we generate the corresponding time-domain
waveform using the \texttt{IMRPhenomPv2\_NRTidal} approximant, with the
sample-specific intrinsic parameters and luminosity distance. The
instantaneous GW energy flux received by an observer in a given direction
is calculated from the two tensor polarizations as
\begin{equation}
F_{\rm GW}(t,\theta,\phi)
=
\frac{c^3}{16\pi G}
\left[
\dot{h}_{+}^{\,2}(t,\theta,\phi)
+
\dot{h}_{\times}^{\,2}(t,\theta,\phi)
\right],
\label{eq:app_gw_flux}
\end{equation}
where overdots denote derivatives with respect to observer-frame time.
The derivatives are evaluated numerically from the generated
time-domain waveforms.

Since gravitational-wave emission is anisotropic, the energy flux
measured along a single viewing direction cannot be extrapolated
isotropically to obtain the total radiated energy. Instead, we calculate
the angle-integrated GW luminosity by decomposing the complex strain
into spin-weighted spherical harmonics. For the dominant quadrupolar
radiation considered here, the strain is expressed as
\begin{equation}
h_+(t,\theta,\phi)-ih_\times(t,\theta,\phi)
=
\sum_{m=-2}^{2}
h_{2m}(t)\,
{}_{-2}Y_{2m}(\theta,\phi),
\label{eq:app_gw_mode_expansion}
\end{equation}
where $h_{2m}(t)$ are the quadrupolar mode amplitudes and
${}_{-2}Y_{2m}(\theta,\phi)$ are the spin-weighted spherical harmonics. 

We employ the spin-weighted spherical-harmonic formalism
\citep{10.1063/1.1705135}, together with the precessing waveform
framework underlying IMRPhenomP \citep{PhysRevLett.113.151101}, to reconstruct
the quadrupolar waveform modes. Since the waveform generator
provides the two gravitational-wave polarizations for a specified
observer direction, we implement a numerical reconstruction of
the five quadrupolar modes using five independent viewing
directions of the same intrinsic binary configuration.

For each posterior sample, the five quadrupolar mode amplitudes
$h_{2m}$ ($m=-2,-1,0,1,2$) are reconstructed using the
\texttt{IMRPhenomPv2\_NRTidal} approximant through
\texttt{LALSimulation} \citep{lalsuite, Wette2020}. The source parameters are taken from
the corresponding low-spin or high-spin Bayesian
parameter-estimation analysis. The binary parameters are first
mapped to the intrinsic parameters required by the
\texttt{SimIMRPhenomP} waveform generator. The intrinsic binary
configuration, including the component masses, spin parameters,
spin orientations, tidal deformabilities, and reference
frequency, is kept fixed throughout the reconstruction. Only
the observer viewing direction is varied.

Specifically, we evaluate the waveform at five viewing angles
whose polar coordinates satisfy
\begin{equation}
\cos\theta_j =
\left\{
0.90,\ 0.45,\ 0,\ -0.45,\ -0.90
\right\},
\qquad j=1,\ldots,5,
\label{eq:five_view_angles}
\end{equation}
with the azimuthal coordinate fixed at $\phi=0$. These five
evaluations therefore correspond to different viewing directions
of the same binary system, rather than independent binary
configurations. A fresh \texttt{LALDict} is initialized for each
waveform evaluation to preserve the required tidal-parameter
ordering.

At each positive Fourier frequency, the complex strain
corresponding to the $j$-th viewing direction is expressed as
\begin{equation}
\tilde{b}_j(f)
=
\tilde{h}_{+}(f;\theta_j)
-i\tilde{h}_{\times}(f;\theta_j)
=
\sum_{m=-2}^{2}
{}_{-2}Y_{2m}(\theta_j,0)
\tilde{h}_{2m}(f).
\label{eq:five_view_decomposition}
\end{equation}
The five viewing directions thus provide five independent
linear equations for the five complex quadrupolar mode
amplitudes at each frequency. Defining the harmonic matrix
\begin{equation}
Y_{jm}={}_{-2}Y_{2m}(\theta_j,0),
\end{equation}
the system can be written in matrix form as
\begin{equation}
\boldsymbol{\tilde{b}}(f)
=
\mathbf{Y}\boldsymbol{\tilde{h}}(f),
\end{equation}
where $\boldsymbol{\tilde{b}}$ contains the five complex
waveform projections and $\boldsymbol{\tilde{h}}$ contains
the five quadrupolar mode amplitudes. The harmonic matrix
constructed using the adopted viewing angles is verified
to be nonsingular. The modes are consequently recovered
at each positive frequency through
\begin{equation}
\boxed{
\boldsymbol{\tilde{h}}(f)
=
\mathbf{Y}^{-1}\boldsymbol{\tilde{b}}(f)
}.
\label{eq:five_view_mode_inversion}
\end{equation}

The negative-frequency components are reconstructed
independently, using the relation
\begin{equation}
\tilde{b}_j(-f)
=
\left[
\tilde{h}_{+}(f;\theta_j)
+i\tilde{h}_{\times}(f;\theta_j)
\right]^*,
\qquad f>0,
\end{equation}
and solving the corresponding harmonic system. This
treatment avoids imposing an inappropriate Hermitian
conjugate-symmetry condition on the individual complex
quadrupolar modes.

The resulting two-sided Fourier-domain modes are then
transformed into the time domain. Their time derivatives
are evaluated spectrally according to
\begin{equation}
\widetilde{\dot{h}}_{2m}(f)
=
2\pi i f\,\tilde{h}_{2m}(f),
\label{eq:gw_mode_derivative}
\end{equation}
before restricting the signals to the adopted
energy-integration interval.

This reconstruction provides the time-dependent
quadrupolar mode amplitudes of the precessing waveform,
thereby retaining the evolving angular structure of the
radiation. The total GW energy can consequently be
calculated by integrating the reconstructed mode power
over the full solid angle, without adopting a fixed
inclination-dependent correction to the waveform observed
along the line of sight.

The reconstruction is independently validated by
comparing the reconstructed waveform projections with
direct waveform evaluations at additional viewing
directions. The resulting relative reconstruction errors
are of order $10^{-15}$ in representative numerical tests,
demonstrating the numerical consistency of the
five-viewing-angle inversion.

\begin{table}[t]
\centering
\caption{Summary of the independent Bayesian parameter-estimation setup
used for GW170817. The posterior samples from both the low- and high-spin
analyses are used in the subsequent calculation of the gravitational-wave
energy.}
\label{tab:gw_pe_setup}
\begin{tabular}{ll}
\hline
\hline
\textbf{Parameter} & \textbf{Analysis setup} \\
\hline
Detectors & LIGO Hanford (H1), LIGO Livingston (L1) \\

Data segment & 128 s centered on merger \\

Merger GPS time & 1187008882.43 \\

Waveform model & \texttt{IMRPhenomPv2\_NRTidal} \\

Inference framework & \texttt{Bilby}/\texttt{parallel\_bilby} \citep{bilby_paper,pbilby_paper}\\

Sampler & \texttt{Dynesty} \citep{dynesty_paper}\\

Chirp mass &
$\mathcal{M}_c \sim \mathcal{U}(1.184,\,2.168)\,M_\odot$ \\

Mass ratio &
$q \sim \mathcal{U}(0.125,\,1)$ \\

Tidal deformabilities &
$\Lambda_1,\Lambda_2 \sim \mathcal{U}(0,\,5000)$ \\

Low-spin prior &
$a_1,a_2 \sim \mathcal{U}(0,\,0.05)$ \\

High-spin prior &
$a_1,a_2 \sim \mathcal{U}(0,\,0.89)$ \\

Spin orientation & Precessing spin orientations \\

Sky position & Fixed to NGC~4993 \\

Luminosity distance &
\shortstack[l]{
Truncated Gaussian: $\mu=40.7\,\mathrm{Mpc}$,
$\sigma=2.4\,\mathrm{Mpc}$,
$d_L\in[35,46]\,\mathrm{Mpc}$
} \\

Binary orientation &
$\cos\theta_{\rm jn}\sim\mathcal{U}(-1,\,1)$ \\

\hline
\hline
\end{tabular}
\end{table}

The total GW luminosity is obtained by integrating the energy flux over the full solid angle,
\begin{equation}
L_{\rm GW}(t)
=
d_L^2
\int_{4\pi}
F_{\rm GW}(t,\theta,\phi)\,d\Omega .
\label{eq:app_gw_luminosity_integral}
\end{equation}
Using the orthonormality of the spin-weighted spherical harmonics,
this expression reduces to
\begin{equation}
L_{\rm GW}(t)
=
\frac{c^3d_L^2}{16\pi G}
\sum_{m=-2}^{2}
\left|\dot{h}_{2m}(t)\right|^2,
\label{eq:app_gw_luminosity}
\end{equation}
where $d_L$ is taken from the corresponding posterior sample. This
angular integration accounts for the anisotropic emission pattern of
the quadrupolar radiation and yields the total luminosity rather than
an isotropic-equivalent estimate along the observer's line of sight.

The source-frame radiated GW energy is then obtained by integrating
the luminosity over the modeled waveform duration and applying the
cosmological redshift correction:
\begin{equation}
\begin{aligned}
E_{\rm GW}
&=
\frac{1}{1+z}
\int_{t_{\rm start}}^{t_{\rm end}}
L_{\rm GW}(t)\,dt
\\
&=
\frac{c^3d_L^2}{16\pi G(1+z)}
\sum_{m=-2}^{2}
\int_{t_{\rm start}}^{t_{\rm end}}
\left|\dot{h}_{2m}(t)\right|^2\,dt ,
\end{aligned}
\label{eq:app_egw}
\end{equation}
where $z$ is the redshift associated with the corresponding posterior
sample. The integration is performed in observer-frame time, with
$t_{\rm start}=t_{\rm ref}=-10$~s. The endpoint $t_{\rm end}$ is
determined separately for each waveform from the numerical saturation
of the cumulative modeled GW energy following its amplitude peak.
The endpoint selection requires a minimum interval of 10~ms after the
waveform-amplitude peak and verifies that the remaining contribution
to the integrated energy is negligible.

This procedure includes the late-time contribution of the generated
waveform but does not independently model the physical post-merger
emission from the remnant. The resulting sample-by-sample estimates
of $E_{\rm GW}$ propagate the uncertainties in the binary parameters,
luminosity distance, and redshift into the GW-energy posterior
distribution. The adopted posterior treatment for the low-spin and
high-spin analyses is described in Section~\ref{GW}. The resulting
GW-energy distribution is used as the gravitational-wave contribution
to the ADM-anchored calorimetric energy budget. A summary of the Bayesian parameter-estimation setup used for the
analysis of GW170817 is presented in Table~\ref{tab:gw_pe_setup}.

\section{Prompt-emission Analysis}
\label{app:prompt}

\subsection{Spectral Analysis}
\label{app:prompt_spectral}

We use the \textit{Fermi}/GBM TTE data from NaI~1, NaI~2, and NaI~5,
covering $\sim8$--$900$~keV, together with BGO~0, covering
$\sim250$~keV--$30$~MeV. The light curves and the selected time interval,
$t=-0.258$ to $+3.328$~s relative to the GBM trigger, are shown in Figures~\ref{fig:grb170817a_spectrum}a and \ref{fig:grb170817a_spectrum}b. The time-integrated spectrum over
this interval is well described by an empirical cutoff power-law (CPL) model.
The best-fit spectral peak energy is
$E_{\rm peak}=51.5^{+12.8}_{-10.6}$~keV and the low-energy index is
$\alpha=1.2\pm0.5$. 
The count spectrum with residuals and the unfolded $\nu F_{\nu}$ spectrum
for the adopted CPL fit are shown in Figures~\ref{fig:grb170817a_spectrum}c and \ref{fig:grb170817a_spectrum}d.

\subsection{Structured-jet Correction}
\label{app_jet_struc}
We account for the off-axis viewing geometry using Gaussian and power-law
structured-jet profiles. Here, $\theta$ denotes the angle measured from
the jet axis, with $\theta_{\rm obs}$ representing the viewing angle of the
observer. We adopt a viewing-angle range
$\theta_{\rm obs}=15^\circ$--$25^\circ$
\citep{Mooley2018,Ghirlanda2019} and a
jet-core ratio $\theta_{\rm obs}/\theta_c=3.5$--$6.5$,
consistent with structured-jet constraints
\citep{Mooley2018,Ghirlanda2019,Troja2019}.

For the Gaussian structure
\citep{GillGranot2018,Troja2019},
the angular isotropic-equivalent energy
is described by
\begin{equation}
E_{\rm iso}(\theta)
=
E_{\rm iso,0}
\exp\left(-\frac{\theta^2}{2\theta_c^2}\right),
\label{eq:gaussian_jet_app}
\end{equation}
where $E_{\rm iso,0}$ is the on-axis isotropic-equivalent energy and
$\theta_c$ is the characteristic core angle. For the power-law structure \citep{BeniaminiGranotGill2020,TakahashiIoka2020},
we adopt
\begin{equation}
E_{\rm iso}(\theta)
=
E_{\rm iso,0}
\left(1+\frac{\theta}{\theta_c}\right)^{-k},
\label{eq:powerlaw_jet_app}
\end{equation}
where $k$ determines the angular decline of the energy profile outside the
core. For each realization, the normalization is constrained by the observed
prompt emission at $\theta=\theta_{\rm obs}$. The corresponding
jet-integrated prompt radiated energy is then obtained by integrating the
angular energy profile over solid angle,
\begin{equation}
E_{\gamma,\rm true}
=
\frac{1}{2}\int_0^\pi
E_{\rm iso}(\theta)\sin\theta\,d\theta ,
\label{eq:prompt_energy_integral}
\end{equation}
for the adopted axisymmetric two-sided jet prescription. The resulting
Gaussian and power-law distributions are used to quantify the systematic
uncertainty associated with the assumed jet angular structure.


\section{Baryonic-Mass Bookkeeping}
\label{app:baryonic_bookkeeping}
Since the GW analysis favours non-zero dimensionless spins ($a_1$ and $a_2$) for both progenitor stars (Figure \ref{fig:gw_posteriors_high_low_spin}a,b), we adopt the rotating neutron-star prescription for the baryonic-gravitational mass conversion. Accordingly, for each gravitational wave posterior sample, the source-frame gravitational
masses of the two progenitor stars are converted to baryonic masses using the
EOS-insensitive baryonic--gravitational mass relation for rotating neutron stars with arbitrary spin periods \citep{2020FrPhy..1524603G},

\begin{equation}
\label{eq:Mb_Mg_app}
M_b
=
M_g
+
0.0777\,M_g^2
+
0.0018\,M_g^3,
\end{equation}
where the masses are expressed in units of $M_\odot$. The total baryonic mass remaining in the bound compact remnant is then obtained by subtracting the baryonic masses associated with the dynamical ejecta and post-merger disk (equation \ref{eq:Mb_rem_main}). 
The resulting $M_{b,\rm rem}$ is converted to the gravitational mass $M_{\rm rem}$ of the immediate post-merger compact remnant by inverting the same relation. The posterior samples are taken from the combined low-spin and high-spin analyses. The central 68\% range of the progenitor baryonic mass obtained from the posterior samples, together with the adopted ranges of the ejecta and disk masses, is propagated by Monte Carlo sampling; for each draw, the relation is solved for $M_{\rm rem}$ and the physically relevant positive solution is retained. The obtained distribution is shown in Figure \ref{fig:remnant_mass_neutrino}a. 

\section{Sensitivity to the Adopted Jet-Energy Fraction}
\label{app:gw190425_fgamma_sensitivity}

To assess the sensitivity of the GW190425 \citep{Abbott2020GW190425,Fletcher2019GBM} inclination constraints on the $d_L - \theta_v$ parameter space due to the GW170817 calibrated prompt energy fraction, we vary the jet-energy fraction $f_{\rm jet}=0.11\%$ adopted from GRB~170817A by a factor of two ($0.055$--$0.22\%$), keeping the prompt radiative efficiency fixed at $0.15$ \citep{Beniamini2016, OConnor2024}.

Across the GW190425 distance posterior ($d_L\simeq88$--$228$~Mpc \, 90\% range), the three Gaussian curves nearly coincide, with the minimum viewing angle confined to $\simeq17^\circ$--$20^\circ$ and shifting by $\lesssim1.2^\circ$ between $0.5\,f_{\rm jet}$ and $2\,f_{\rm jet}$ (Figure~\ref{fig:gw_posteriors_high_low_spin}c,d). For the $k=5$ power-law jet, the minimum viewing angle spans $\simeq21^\circ$--$43^\circ$, with the $0.5\,f_{\rm jet}$ and $2\,f_{\rm jet}$ curves separated by $\simeq8^\circ$--$11^\circ$. The Gaussian constraint is, therefore, insensitive to the assumed jet energy, whereas the power-law constraint is more sensitive; in both cases, the non-detection remains compatible with an off-axis viewing geometry.

\bibliography{Ref}
\bibliographystyle{aasjournalv7}



\end{document}